\documentclass[prl,aps,superscriptaddress]{revtex4}
\usepackage{bm}
\usepackage{amsfonts}
\usepackage[dvips]{graphicx}
\usepackage{mathrsfs}
\usepackage[intlimits]{amsmath}
\usepackage[colorlinks, citecolor=blue]{hyperref}

\begin{document}

\title{Observation of entanglement sudden death and rebirth \\
by controlling solid-state spin bath}
\author{F. Wang$^{1}$, P.-Y. Hou$^{1}$, Y.-Y. Huang$^{1}$, W.-G. Zhang$^{1}$%
, X.-L. Ouyang$^{1}$, X. Wang$^{1}$, X.-Z. Huang$^{1}$, H.-L. Zhang$^{1}$,
L. He$^{1}$, X.-Y. Chang$^{1}$, L.-M. Duan\thanks{%
To whom correspondence should be addressed. E-mail: lmduan@umich.edu.}}
\affiliation{Center for Quantum Information, IIIS, Tsinghua University, Beijing 100084,
PR China}
\affiliation{Department of Physics, University of Michigan, Ann Arbor, Michigan 48109, USA}

\begin{abstract}
Quantum entanglement, the essential resource for quantum information
processing, has rich dynamics under different environments. Probing
different entanglement dynamics typically requires exquisite control of
complicated system-environment coupling in real experimental systems. Here,
by a simple control of the effective solid-state spin bath in a diamond
sample, we observe rich entanglement dynamics, including the conventional
asymptotic decay as well as the entanglement sudden death, a term coined for
the phenomenon of complete disappearance of entanglement after a short
finite time interval. Furthermore, we observe counter-intuitive entanglement
rebirth after its sudden death in the same diamond sample by tuning an
experimental parameter, demonstrating that we can conveniently control the
non-Markovianity of the system-environment coupling through a natural
experimental knob. Further tuning of this experimental knob can make the
entanglement dynamics completely coherent under the same environmental
coupling. Probing of entanglement dynamics, apart from its fundamental
interest, may find applications in quantum information processing through
control of the environmental coupling.
\end{abstract}

\maketitle

Besides its significance as a fundamental concept in quantum mechanics,
entanglement has been recognized as an essential resource for quantum
computation and communication \cite{1,2,3,4}. In any real experimental
system, due to its inevitable coupling to the surrounding environment,
entanglement degrades with time, leading to various kinds of entanglement
dynamics under different environmental couplings \cite{5,6,7,8,9}. The most
common one is the asymptotic decay of the entanglement, where the
entanglement approaches zero (typically exponentially) as the time goes to
infinity. This behavior is similar to the quantum coherence decay and arises
when the environmental coupling is dominated by pure dephasing
\cite{5,6}. Under more complicated dissipation, entanglement may completely
vanish in a finite (typically short) time interval, a phenomenon called the
entanglement sudden death (ESD) \cite{7,8,9,10,11}. The ESD is identified to
be more disruptive to quantum information processing due to the fast
disappearance of entanglement \cite{7,8,9}. Under more unusual situations
which require non-Markovianity of the system-environment coupling, the
entanglement can reappear after its sudden death for a while, which is
termed as the entanglement rebirth \cite{9}. The entanglement sudden death
and rebirth have been extensively studied theoretically \cite%
{12,14,15,16,17,18,19}. On the experimental side, all-optical quantum
experiments can simulate environmental couplings of photonic qubits by
linear optics elements to demonstrate the ESD \cite{20,21} as well as the
non-Markovian coupling \cite{22,23,24}. ESD is also observed between atomic
ensembles but with no entanglement rebirth or non-Markovian behavior \cite%
{25}. It is desirable to find an experimental system where the natural
system-environment coupling and its Markovianity can be controlled to probe
different kinds of entanglement dynamics in a single system, including the
ESD and the entanglement rebirth. Probe of the non-Markovian dynamics plays an
important role in control of open quantum systems \cite{26,27,28}.

In this paper, we demonstrate that rich entanglement dynamics can be
observed in a single diamond sample by controlling the effective coupling of
spin qubits to the solid-state spin bath. The system-environment coupling is
controlled by the dynamical decoupling pulses, which provide a tunable
filter function to modify the contributing noise spectrum of the bath.
Depending on the control parameter in this filter function, we observe the
Markovian asymptotic decay as well as the sudden death of the entanglement.
Tuning of the same parameter allows us to go to non-Markovian region of the
dissipation, where we observe the entanglement rebirth after the ESD.
Further tuning of this parameter can make the entanglement dynamics
completely coherent under this spin bath, showing entanglement Rabi
oscillations.

Our experimental demonstration makes use of the hybrid spin system composed
of the electron spin and the host $^{14}N$ nuclear spin associated with the
nitrogen-vacancy center in diamond (Fig. 1(a)). The surrounding $^{13}C$
nuclear spins in the crystal lattice act as a natural reservoir and are coupled to
the electron spin via varying anisotropic hyperfine interactions (Fig.
1(a)). The impact of the $^{13}C$ bath on the nitrogen nuclear spin can be
ignored due to the lower interaction strength compared to that of the
electron spin.

\begin{figure}[tbp]
\includegraphics[width=160mm]{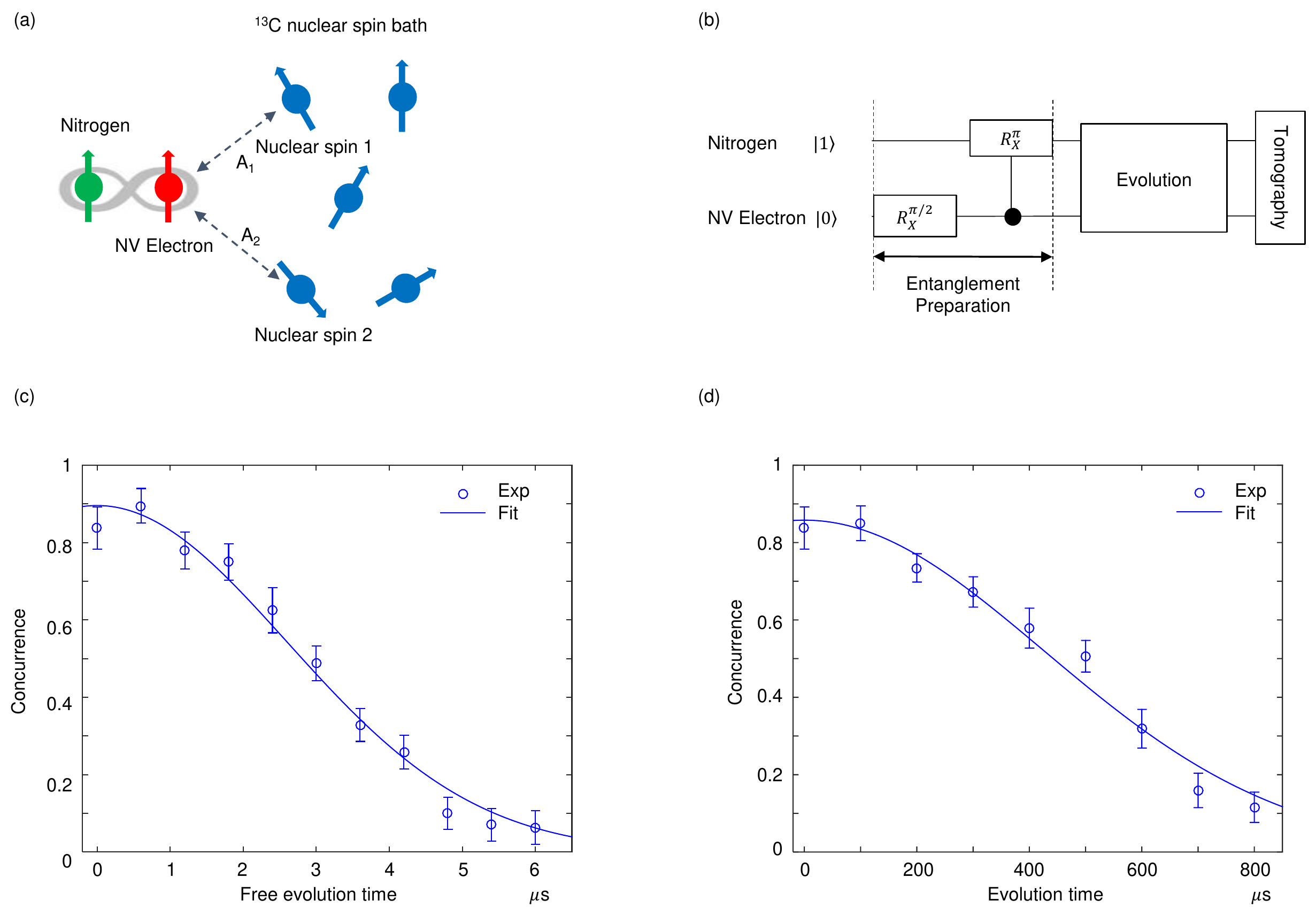}
\caption{Experimental system. (a) The NV electron spin (red), the host
nitrogen nuclear spin (green) and the coupled $^{13}C$ spin bath (Blue).
Entangled state is prepared on the electron-nitrogen pair and is exposed to
the spin bath. Hyperfine parameters of distinct carbons are denoted as $%
\mathbf{A}_i$ for carbon $i$. (b) Gate sequence to prepare entanglement in
the electron-nitrogen nuclear spin pair. The conditional $\protect\pi$ rotation is
implemented by a $2.928$ MHz radio-frequency (rf) signal with two fast $\protect\pi$ rotations
of the electron spin symmetric on both sides to protect the coherence of the
electron spin. (c) Entanglement decay as a function of free evolution time
with the system subject to the fluctuating spin bath. (d) Entanglement decay
as a function of evolution time under the Hahn echo. Solid lines are fits by the function $%
exp(-(t/T_c)^2)$ with $T_c=3.7$ $\mu s$ ($602$ $\mu s$) for Fig. c (Fig. d). The error bars in this and the following
figures denote one standard deviation.}
\end{figure}

The experiments are performed at room temperature on a bulk diamond sample
with a natural $^{13}C$ abundance. Under an external magnetic field of $479$
Gauss along the NV symmetry axis, both the electron and the nitrogen nuclear
spin are polarized via optical pumping \cite{29}. We prepare entangled state
between the two spins using a rf-induced $\pi$ rotation of the nuclear spin
conditional on the state of the electron spin (Fig. 1(b)). Coherence of the
electron spin is protected by two fast $\pi$ rotations symmetric on both
sides of the rf pulse with the form $\tau_{1}-\pi-2\tau_{1}-\pi-\tau_{1}$, where $2\tau_{1}$ is the rf pulse duration.
Final state density matrix is reconstructed from measurements by two-bit state tomography
\cite{30}.

Entanglement for a two-qubit state $\rho$ can be quantified by concurrence C
\cite{31}, which is given by $C=max\{0,\Gamma\}$ where $\Gamma=\sqrt{%
\lambda_1}-\sqrt{\lambda_2}-\sqrt{\lambda_3}-\sqrt{\lambda_4}$ with $%
\lambda_i$ denoting the eigenvalues of the matrix $\rho(\sigma_y\otimes%
\sigma_y)\rho^*(\sigma_y\otimes\sigma_y)$ in decreasing order, and $\rho^*$
denoting the complex conjugate of $\rho$ in the computational basis $%
\{|00\rangle,|01\rangle,|10\rangle,|11\rangle\}$. This quantity can be
interpreted as a measure for the quantum correlation present in the system
and takes the value from $0$ to $1$.

We start by studying the dynamics of entanglement under free evolution of
the two spins. Under this condition, the electron spin is subject to fast
decoherence noise induced by the bath. Therefore, asymptotic decay of
entanglement is observed by fitting the concurrence to $exp(-(t/T_c)^2)$
(Fig. 1(c)), where $T_c$ is the electron spin coherence time. A Hahn echo
can cancel part of the interaction with the spin bath and thus significantly increase the
coherence time. As a result, the observed decay of entanglement under the Hahn echo is also
asymptotic but with a much slower decay rate as shown in Fig. 1(d).

\begin{figure}[tbp]
\includegraphics[width=160mm]{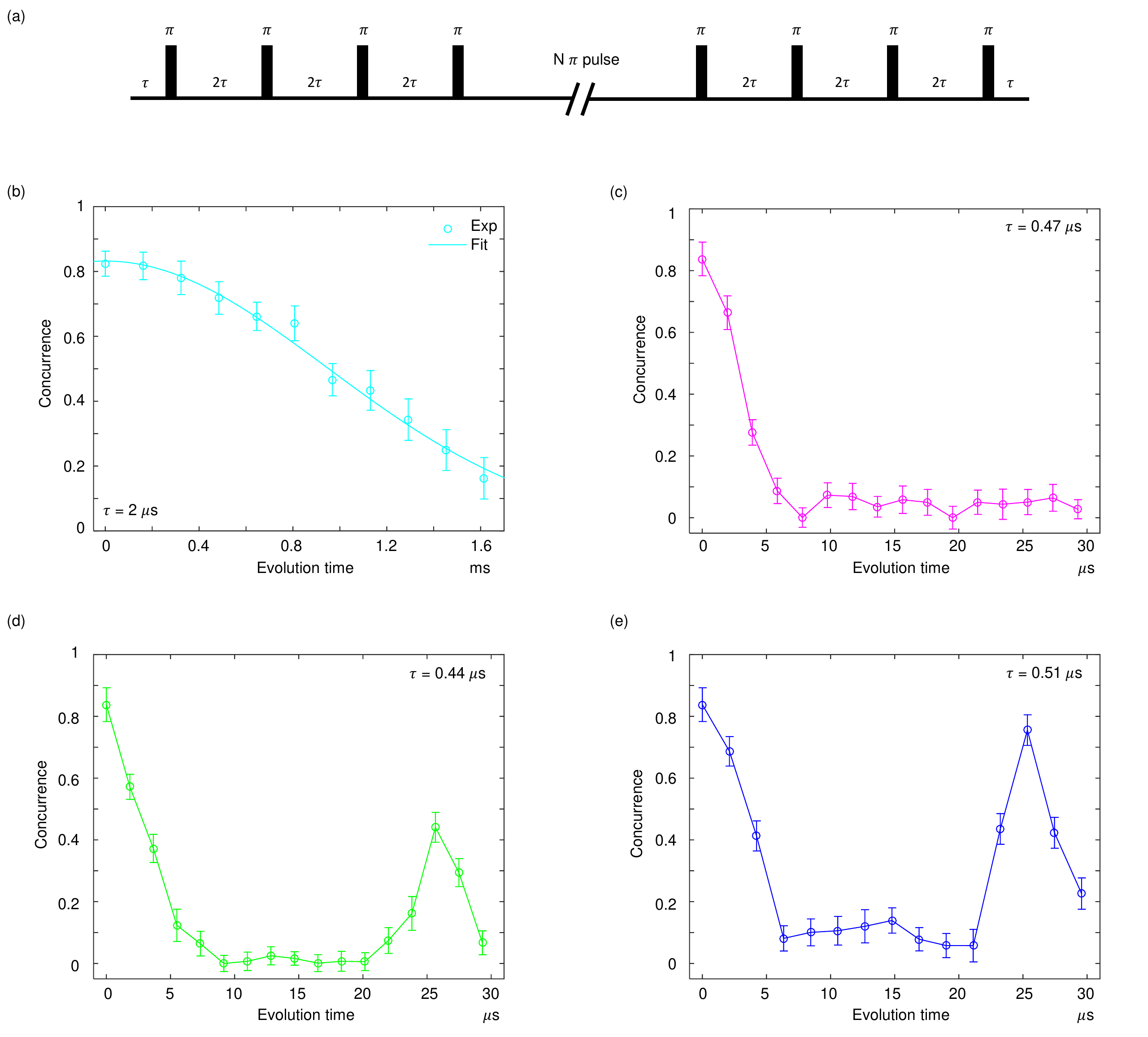}
\caption{Decay of electron-nuclear spin entanglement under CPMG sequences. (a)
Diagram of the CPMG sequence. (b) Asymptotic decay of entanglement as a
function of the total evolution time at $\tau=2$ $\mu s$. Solid
line is a fit by the function $exp(-(t/T_2)^2)$ with $T_2=1.33$ $ms$. (c) Observation of entanglement sudden death as a
function of the evolution time at $\tau=0.47$ $\mu s$.
(d, e) Non-Markovian entanglement dynamics as a function of the total
evolution time at $\tau=0.44$ $\mu s$ and $\tau=0.51$ $\mu s$, which shows entanglement sudden death and rebirth.}
\end{figure}

A more complicated behavior of entanglement dynamics can be observed if one extends
the Hahn echo to periodic repetitions of the Carr-Purcell-Meiboom-Gill
(CPMG) sequence (Fig. 2(a)). The CPMG sequence acts as a filter of the bath
in the frequency domain, whose center frequency can be tuned by changing the
inter-pulse duration $2\tau$. In experiments we selectively apply four
specific filters to the system by choosing corresponding $\tau$ and monitor
the evolution of entanglement by gradually increasing the number of pulses $N
$ (and therefore the total evolution time). Three distinct behaviors of
entanglement dynamics are observed: (i) asymptotic decay with $\tau =2$ $\mu s$ (Fig.
2(b)), (ii) entanglement sudden death with $\tau =0.47$ $\mu s$ (Fig. 2(c)),
(iii) non-Markovian behavior which shows entanglement sudden death and
rebirth with $\tau=0.44$ $\mu s$ and $\tau =0.51$ $\mu s$ (Fig. 2(d,e)).

\begin{figure}[tbp]
\includegraphics[width=160mm]{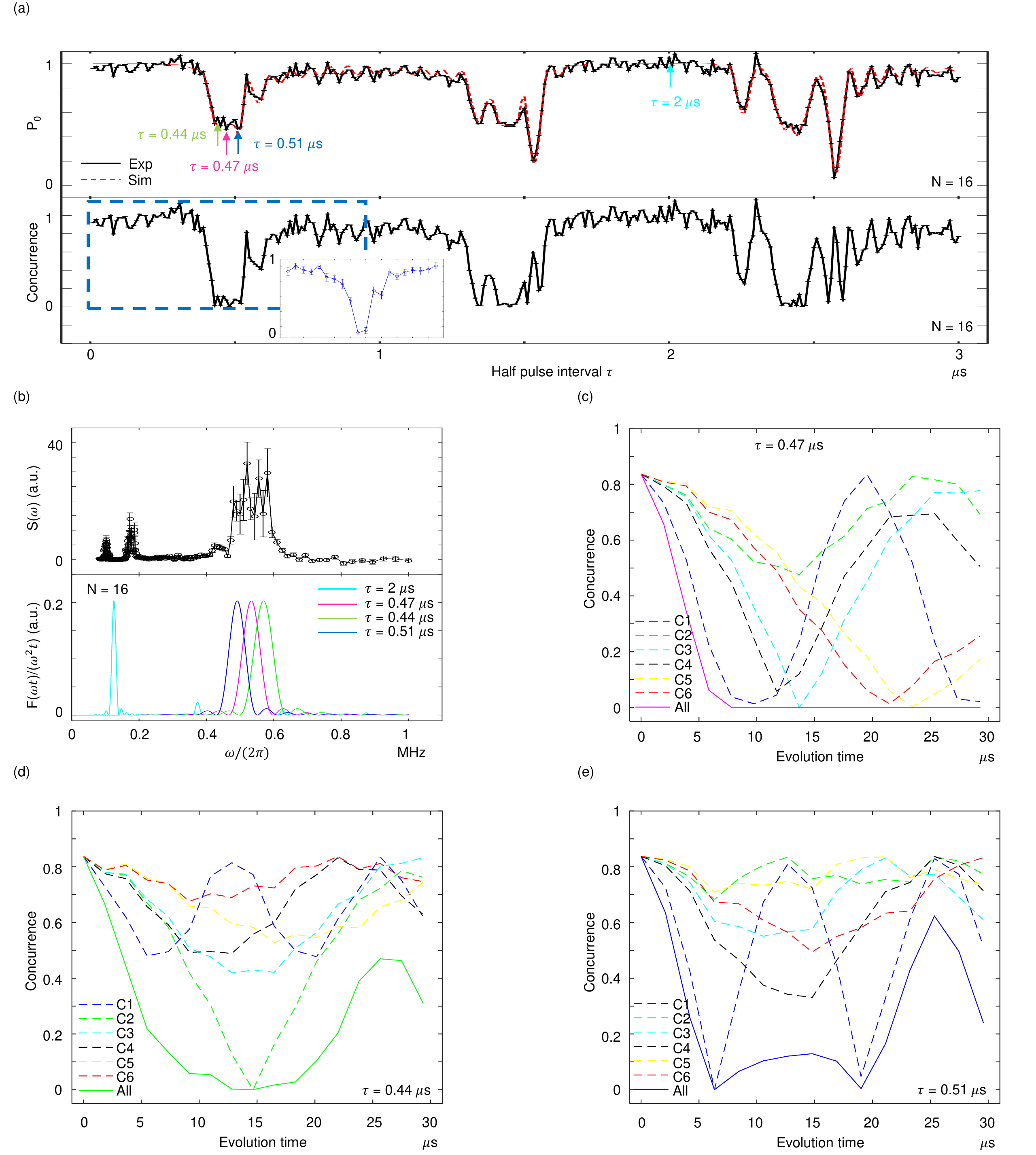}
\caption{Analysis of the entanglement evolution under CPMG sequences. (a)
Upper panel: decay of electron coherence as a function of half inter-pulse
duration $\protect\tau$ with $N=16$. Black dots and solid line are
experimental data taken every $10$ ns. Red dashed line denotes the simulation result
under six identified carbon nuclear spins in the bath with their hyperfine parameters calibrated by the method
described in the supplementary material \cite{41}. Arrows indicate corresponding $%
\protect\tau$ in Fig. 2(b,c,d,e) respectively. Lower panel: calculated
entanglement concurrence from electron coherence in the upper figure. Inset:
experimental results with the same time range in the blue dashed square.
Blue circles are experimental data taken every $50$ ns. (b) Upper panel:
noise spectrum constructed from electron spin coherence in Fig. 3(a).
Central components correspond to the first order resonance of the nuclear spin
bath. Higher order resonance terms are shown as peaks at lower frequencies.
Lower panel: filter function of the CPMG sequences with $N=16$. Peaks appear
at $\protect\omega_0/(2\protect\pi)=1/(4\protect\tau)$. (c,d,e) Simulation
of entanglement decay under the same condition described in Fig. 2(c,d,e)
respectively. Dashed lines are results corresponds to single identified
carbon nuclear spin. Solid line is the result with effects of six carbon nuclear spins
combined together.}
\end{figure}

The phenomena can be explained by the spectral filtering of the spin bath through the CPMG sequence, which
either suppresses \cite{32,33,34} or resonantly amplifies \cite{35} the interactions of the
electron spin with the structured bath. The power spectrum $S(\omega)$ of
the bath includes two parts \cite{36}: thermal noise due to the random
orientations and flip-flops of the bath spins at room temperature, and
dynamical quantum noise caused by the evolution of nuclear spins. Under the CPMG
sequence, both components of the surrounding bath can be detected by
measuring the coherence decay $W(t)$ of the electron spin \cite{37,38}
\begin{equation}
W(t)=exp[-\chi(t)]=exp\left[-\int_{0}^{\infty}\frac{d\omega}{\pi}\frac{S(\omega)}{%
\omega^2}F_N(\omega t)\right]
\end{equation}
where
\begin{equation}
F_N(\omega t)=8sin^2(\frac{\omega t}{2})\frac{sin^4(\frac{\omega t}{4N})}{%
cos^2(\frac{\omega t}{2 N})}
\end{equation}
is the filter function \cite{39} associated with the pulse number $N$ and the total
evolution time $t=2N\tau$ of the CPMG sequence. The detected probability
corresponds to $P_0=(1/2)(W(t)+1)$.

We focus on the dynamical quantum noise of the bath, as the effects of
static thermal noise are removed by the CPMG sequence. According to equation (1),
the precession of the $^{13}C$ spin bath produces a coherence dip of
electron spin centered at the nuclear Larmor frequency $\omega_L=\gamma_C B_z
$, where $\gamma_C$ is the nuclear spin gyromagnetic ratio and $B_z$ is the
external magnetic field. Due to the frequency shift caused by the hyperfine
interactions of multiple nuclear spins in the bath, the observed signal is
shown as a broad collapse instead of a dip (upper panel of Fig. 3(a)).
Higher order resonances of the bath are observed as broadening of the
collapses and splitting of isolated dips. The influence of the bath on the
electron spin coherence takes effect on the entangled state shared between
the electron and the nuclear spin and leads to disruption of entanglement as
shown in lower panel of Fig. 3(a), where entanglement concurrence is
calculated from the detected $P_0$ in the upper panel. We experimentally
measure the entanglement concurrence around the bath and observe a
consistent behavior with the calculation prediction (Inset of Fig. 3(a)).

Spectrum of the dynamical quantum noise can be constructed from the
coherence decay of the electron spin under the CPMG sequence with a spectrum
decomposition method \cite{37}. As shown in the upper panel of Fig. 3(b),
central components in the spectrum correspond to first order resonance of
the nuclear spin bath and peaks in the lower frequencies are higher order
resonance terms. With fixed inter-pulse duration $2\tau$, filter function of the
CPMG sequence covers a narrow frequency region centered at $%
\omega_0=\pi/(2\tau)$ (See lower panel of Fig. 3(b) for filter function). If
we control the filter peak to be off resonant from the bath (for example $%
\tau=2\mu s$), bath influence on the electron spin is sufficiently
suppressed, thus concurrence of entanglement is shown as Markovian
asymptotic decay and can be kept to the limit of electron spin $T_2$ (Fig. 2(b)).
On the contrary, if we control the filter to be on resonant with the bath,
interactions of the electron spin with multiple carbons are amplified,
leading to information flow between the electron spin and the bath. To
understand the mechanism of the information flow, we experimentally identify
six carbon nuclear spins in the bath which give strong influence on the electron spin coherence \cite{40}  (Fig. 3(a) and
Fig. 1 in \cite{41}) and perform simulations with the $8$-qubit system \cite{41}. From
Fig. 3(c), when the filter amplifies the center of the bath ($\tau=0.47\mu s$%
), information exchanges between the electron spin and several carbons with
separate exchange frequency are mixed together and on average are shown as
dissipation from the electron spin to the bath. Consequently, entanglement
vanishes completely in a finite time and ESD is observed. Comparatively,
when the filter amplifies the edge of the bath ($\tau=0.44\mu s$ and $%
\tau=0.51\mu s$), information flow is dominant by the interaction of the
electron spin with the resonant carbon (C1 with $\tau=0.44\mu s$, C2 with $%
\tau=0.51\mu s$), thus revival of entanglement is observed after the
concurrence drops to the minimal value (Fig. 3(d,e)). Full revival is
prohibited by the interactions with other carbons in the bath.

\begin{figure}[tbp]
\includegraphics[width=160mm]{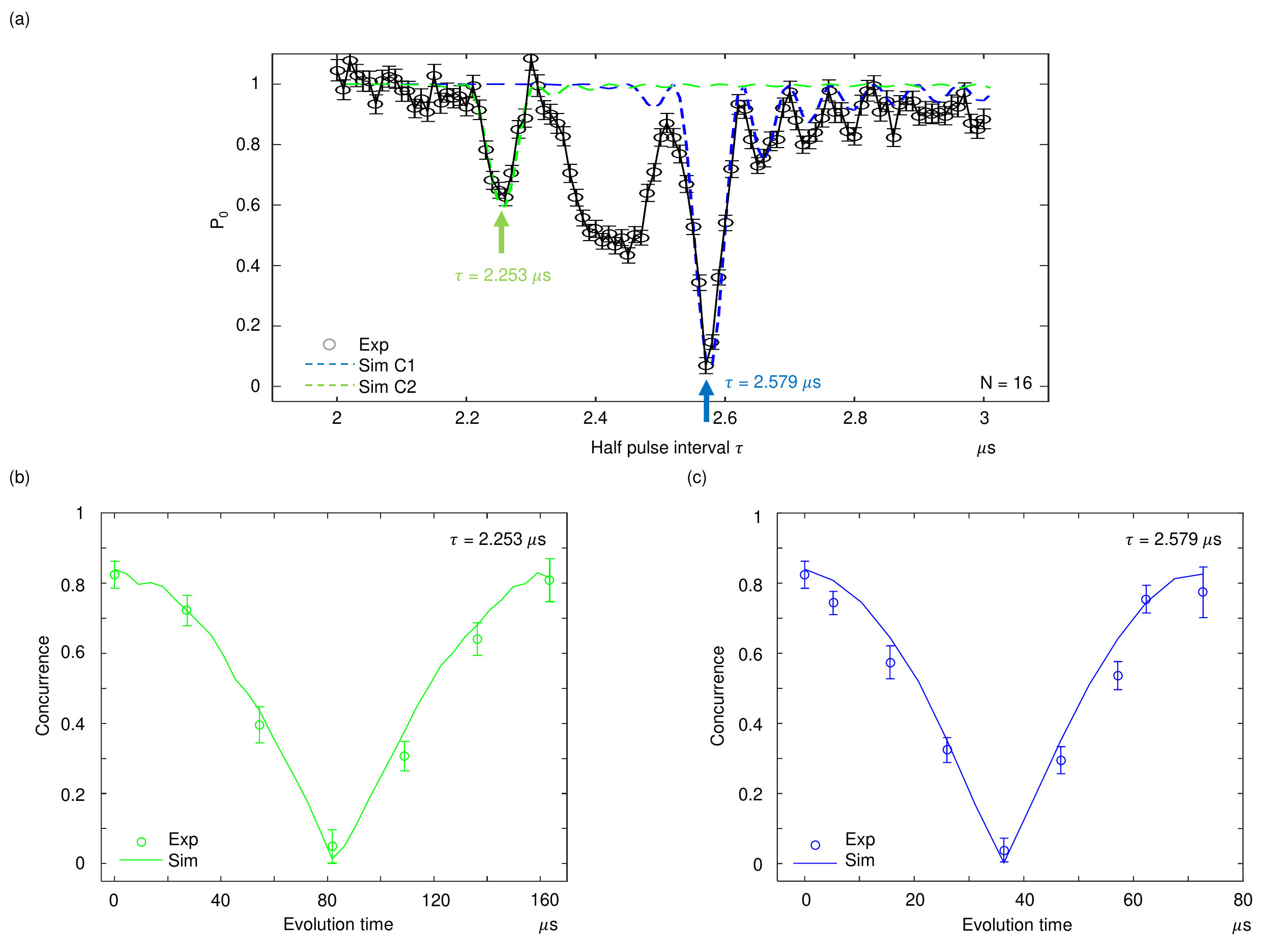}
\caption{Decay of electron coherence and electron-nuclear spin entanglement under
the CPMG sequences with inter-pulse duration $\protect\tau$ on resonant with a
single carbon. (a) Decay of electron coherence as a function of half
inter-pulse duration $\protect\tau$ under CPMG sequences with $N=16$. Black
circles are experimental data taken every $10$ns. (b) Entanglement decay as
a function of the total evolution time with the half inter-pulse duration $%
\tau=2.253$ $\mu s$. The sequence is on resonant with carbon 2
(green arrow in Fig. 4(a)). (c) Entanglement decay as a function of the
total evolution time with the half inter-pulse duration $\tau=2.579$
$\mu s$. The sequence is on resonant with carbon 1 (blue arrow in
Fig. 4(a)). Blue (green) solid lines are simulation results with calibrated
parameters for carbon 1 (2).}
\end{figure}

For the initially entangled states evolving under the CPMG sequence, concurrence
coincides with the measure for quantum non-Markovianity of the electron-bath
system \cite{22}, which in the standard approach is defined by the trace
difference between two quantum states $D(\rho_1(t),\rho_2(t)=(1/2)||%
\rho_1(t)-\rho_2(t)||$, where $\rho_{1,2}(t)=|\phi_{1,2}(t)\rangle\langle
\phi_{1,2}(t)|$ and $|\phi_{1,2}(0)\rangle=(1/\sqrt{2})(|0\rangle_e\pm|1%
\rangle_e)$. Therefore, the observed revival of entanglement also quantifies
the memory effect of the system.

To further probe the non-Markovianity of the system, we control the filter
to be on resonant with isolated single carbon where influence of the
unwanted spin bath is suppressed (Fig. 4(a)). Under this condition, the
isolated carbon acts as a register in a way that coherence information flows
back and forth between the electron-nuclear pair and the electron-carbon
pair, thus leads to entanglement Rabi oscillations. As can be observed from
Fig. 4(b,c), after dropping to $0$, the concurrence immediately begins to
grow and recovers to the initial value after a half cycle. The period of this
entanglement Rabi oscillation is dependent on the transverse component of the hyperfine
parameters ($A_{xz 1}=114.5(1)kHz$, $A_{xz 2}=58.7(3)kHz$).

In summary, beyond demonstrating a perspective insight into the entanglement
dynamics, our observations provide a possible method to control entanglement
in a hybrid system coupled to a natural reservoir. Moreover, measurement for
quantum non-Markovianity allows for the study of memory effects emerging
from correlations within the environment.

\end{document}


\title{Supplementary online material for \\Observation of entanglement sudden death and rebirth \\ by controlling
		solid-state spin bath}
	
	\author{F. Wang$^{1}$, P.-Y. Hou$^{1}$,
		Y.-Y. Huang$^{1}$, W.-G. Zhang$^{1}$, X.-L.
		Ouyang$^{1}$, X. Wang$^{1}$, X.-Z. Huang$^{1}$, H.-L. Zhang$^{1}$, L. He$^{1}$,
		X.-Y. Chang$^{1}$, L.-M. Duan\footnote{To whom correspondence should be addressed. E-mail:
			lmduan@umich.edu.}}
	\affiliation{Center for Quantum Information, IIIS, Tsinghua University, Beijing 100084,
		PR China}
	\affiliation{Department of Physics, University of Michigan, Ann Arbor, Michigan 48109, USA}

	\maketitle

	\section{Experimental setup and state detection}
	
	We use a home-built confocal microscopy system to optically address single NV
	center in the diamond. Initialization and readout of electron spin associated with single NV center is achieved by a $532$ nm green laser controlled by two acoustic optical modulators (AOMs). Both the AOMs are set in a double pass configuration to control the on-off and constrain the leakage of the green laser. After optical mode shaping through a single mode fiber, the green laser is reflected by a wave length dependent Dichroic Mirror (DM) and focused by an
	oil immersed objective lens onto the bulk diamond sample. The fluorescent photons pass through the same DM mirror and get collected by a single photon detector with a $637$ nm long pass filter.
	
	The sample is attached to a cover glass which is mounted on a $3$-axis closed-loop piezo. The microwave signal used for manipulation of electron spin is generated by a carrier signal modulated at an IQ mixer by two analog outputs of an Arbitrary Waveform Generator (AWG) and delivered with a waveguide transmission line fabricated on the cover glass. The radio frequency signal for manipulation of nitrogen nuclear spin is generated by another analog channel of the AWG and delivered with a fabricated coplanar coil on a PCB board, which is mounted and detached from the other side of the sample. Both signals are amplified before sent into the sample to achieve reasonable rabi frequencies.
	
	All the experiments are performed at room temperature with at least $10^6$ repetitions for measurement of each data point. The electron spin state is calculated with $P_0 = (C_{Signal}-C_{Dark})/(C_{Bright}-C_{Dark})$ in each experiment cycle, where $P_0$ is the probability for state $|m_s=0\rangle$, $C_{Signal}$ is number of photons collected for $300$ ns right after the detection laser arises, $C_{Bright}$ and $C_{Dark}$ are number of photons collected for the same period of time with electron spin at $|m_s=0\rangle$ and $|m_s=-1\rangle$ states separately.
	
	Entanglement measurements are performed by mapping the two-bit information onto the electron spin population as described in Ref \cite{1}. Final state density matrix for the entangled state is extracted from the two-bit tomography result with a maximum likelihood calculation.

	\section{Numerical simulation}
	
	The numerical simulation is performed in the rotating frame in the eight-qubit system composed of electron spin, nitrogen nuclear spin and six weakly coupled $^{13}C$ nuclear spins with the Hamiltonian
	\begin{equation*}
	H=\Delta S_z^2+\gamma_e B_z S_z+\gamma_n B_z I_{nz}+A_{\parallel} S_zI_{nz}+Q I_{nz}^2+\sum_{k=1}^{6}{\gamma_c B_z I_{cz}^{(k)}}+S_z\sum_{k=1}^{6}{A_{zz}^{(k)}I_{cz}^{(k)}}+S_z\sum_{k=1}^{6}{A_{xz}^{(k)}I_{cx}^{(k)}}
	\end{equation*}
	Here, $S_z$ and $I_{nz}$ are the spin-$1$ electron and nitrogen nuclear spin operator and $I_{cx(cz)}^{(k)}$ is the spin-$1/2$ nuclear spin operator for the $kth$ carbon. Also, $\Delta=2.87$ $GHz$ denotes the zero field splitting of the electron spin and $Q=-4.945$ $MHz$ denotes the quadrupolar interaction of the nitrogen nuclear spin. Gyromagnetic ratios of electron, nitrogen and carbon nuclear spin are represented with $\gamma_e=2.8$ $MHz/G$, $\gamma_n=-0.308$ $kHz/G$ and $\gamma_c=-1.07$ $kHz/G$ respectively. Hyperfine interactions are represented with $A_{\parallel}=-2.162MHz$ for parallel component of the nitrogen nuclear spin and $A_{zz(xz)}^{(k)}$ for the parallel (transverse) component of the $kth$ carbon nuclear spin. Magnetic field is set at $B_z=479$ Gauss. We assume no interactions between the nuclear spins and a noiseless environment without decoherence and relaxation.
	
	The hyperfine parameters of weakly coupled carbons are first calibrated by fitting the experimental data on the measured electronic spin coherence after the CPMG sequence to the numerical simulation of the corresponding dynamics with the fitting parameters $A_{zz}$ and $A_{xz}$. Afterwards, carbon 1 and 2 are further calibrated using the nuclear ODMR based method described in Ref \cite{2} to a resolution of $0.05$ $kHz$ in $A_{zz}$ and $0.5$ $kHz$ in $A_{xz}$. Hyperfine parameters of other carbons remain at a lower resolution because of the relative long time (compared to the electron coherence time) to initialize the target nuclear spin evolved in the nuclear ODMR method. See Table 1 for the calibrated hyperfine parameters. The resolved six carbons in the bath have the dominant influence on the electron spin coherence dynamics (Fig. 1). Influence of other carbons with smaller $A_{xz}$ are ignored in the simulation, and they have small influence on the experimental data.
	
	\begin{table}[!htb]
		\begin{center}
			\begin{tabular}{ c | c | c }
				Carbon & $A_{zz}(kHz)$ & $A_{xz}(kHz)$ \\ \hline
				1 & -77.02(3) & 114.5(1) \\ \hline
				2 & 71.03(3) & 58.7(3) \\ \hline
				3 & 4(1) & 57(7) \\ \hline
				4 & -13.9(8) & 65(4) \\ \hline
				5 & 16(5) & 37(9) \\ \hline
				6 & -20(3) & 41(10)\\
			\end{tabular}
			\caption{Hyperfine parameters for the six carbons. $A_{zz}$ is the parallel component and $A_{xz}$ is the transverse component. The number in the parenthesis denotes uncertainty in the last digit.}
		\end{center}
	\end{table}
	
	\begin{figure}[tbp]
		\includegraphics[width=160mm]{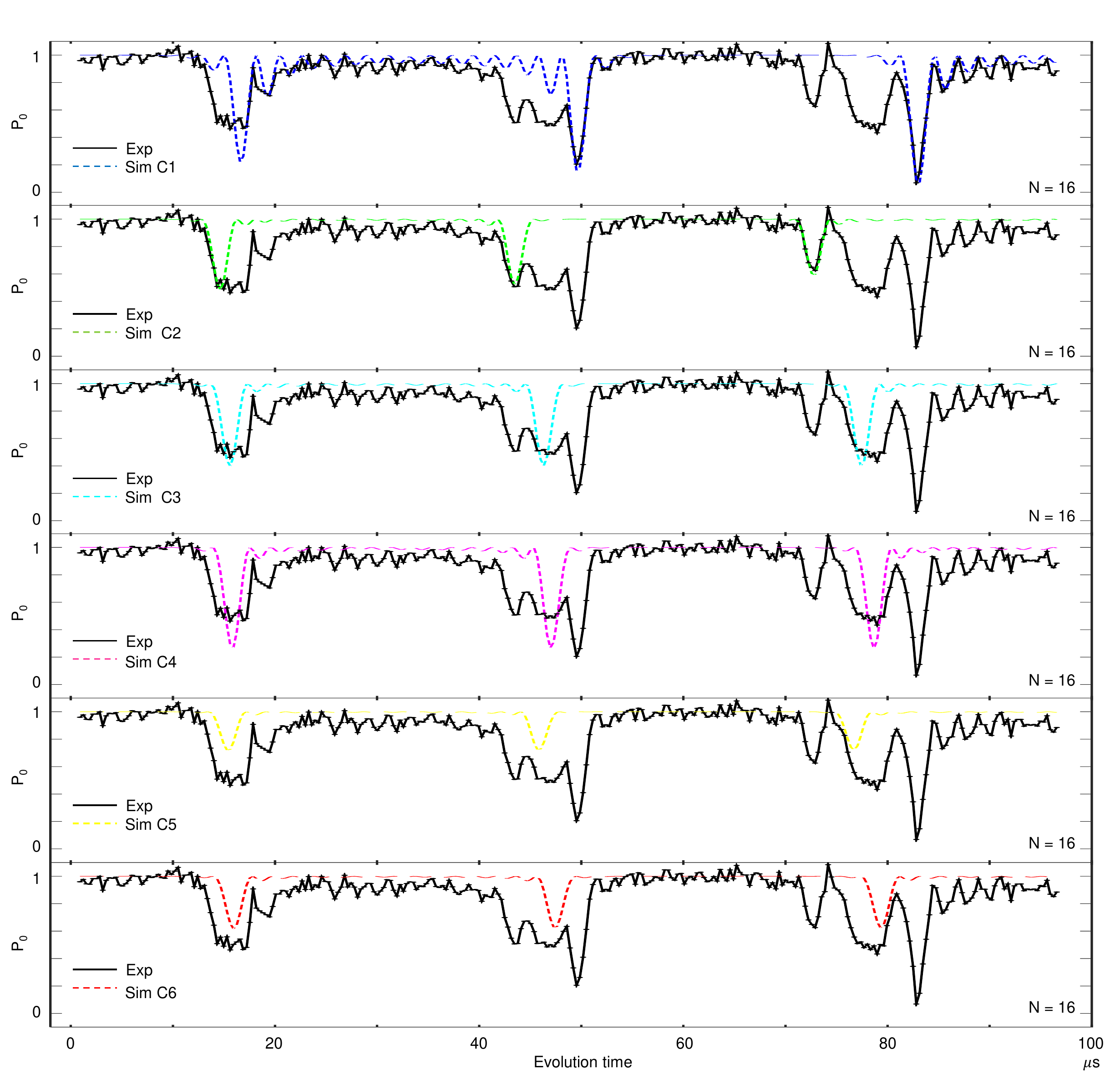}
		\caption{Coherence of electron spin under CPMG sequences. Black dots and lines are experimental data taken every $10$ ns. Dashed lines are simulation results with single carbons respectively.}
	\end{figure}